\documentclass{article}

\usepackage{arxiv}

\usepackage[utf8]{inputenc} 
\usepackage[T1]{fontenc}    
\usepackage{hyperref}       
\usepackage{url}            
\usepackage{booktabs}       
\usepackage{amsfonts}       
\usepackage{nicefrac}       
\usepackage{microtype}      
\usepackage{lipsum}
\usepackage{graphicx}
\graphicspath{ {./images/} }
\usepackage{natbib}
\usepackage{threeparttable}

\title{U-Det: A Modified U-Net architecture with bidirectional feature network for lung nodule segmentation}

\author{
 Nikhil Varma Keetha \\
  Department of Physics\\
  Indian Institute of Technology (Indian School of Mines) Dhanbad\\
  Jharkand - 826004, India \\
  \texttt{keethanikhil@gmail.com} \\
   \And
 Samson Anosh Babu P \\
  Department of CSE\\
  Indian Institute of Technology (Indian School of Mines) Dhanbad\\
  Jharkand - 826004, India \\
  \texttt{samson.enosh@gmail.com} \\
  \And
 Chandra Sekhara Rao Annavarapu \\
  Department of CSE\\
  Indian Institute of Technology (Indian School of Mines) Dhanbad\\
  Jharkand - 826004, India \\
  \texttt{acsrao@iitism.ac.in} \\
}

\begin{document}
\maketitle
\begin{abstract}
Early diagnosis and analysis of lung cancer involve a precise and efficient lung nodule segmentation in computed tomography (CT) images. However, the anonymous shapes, visual features, and surroundings of the nodule in the CT image pose a challenging problem to the robust segmentation of the lung nodules. This article proposes U-Det, a resource-efficient model architecture, which is an end to end deep learning approach to solve the task at hand. It incorporates a Bi-FPN (bidirectional feature network) between the encoder and decoder. Furthermore, it uses Mish activation function and class weights of masks to enhance segmentation efficiency. The proposed model is extensively trained and evaluated on the publicly available LUNA-16 dataset consisting of 1186 lung nodules. The U-Det architecture outperforms the existing U-Net model with the Dice similarity coefficient (DSC) of  82.82\% and achieves results comparable to human experts. 
\end{abstract}


\section{Introduction}
\label{sec1}
Most of the prominent cancer deaths are due to Lung Cancer, and it has a relatively low five-year survival rate of 18 \% \citep{siegel2016cancer}. One of the primary causes of these pulmonary nodules formation is irregular and uncontrollable growth of cells in the lung parenchyma. Detection and analysis of these nodules in the lung tissue in an early phase drastically improve the chance of survival of the patient and facilitates efficient treatment \citep{el2011lung}. 

\begin{figure*}[!ht]
\centering
\includegraphics[width=17cm,height=3.0cm]{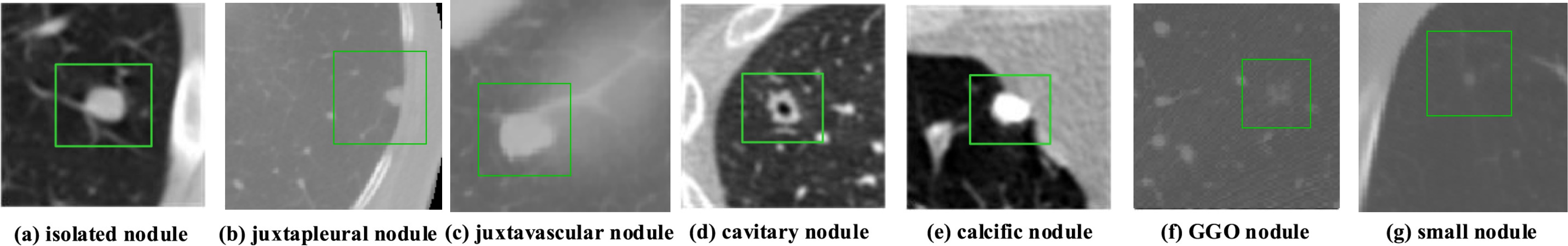}
\caption{Illustrations of various types of lung nodules present in the CT scans. In (g), small nodule indicates a nodule of diameter $ \leq $ \textbf{4 mm}.}
\label{fig1}
\end{figure*}

Computed tomography (CT) scans are a widely used and highly accurate format for screening and analysis of lung nodules. These scans are obtained using multi-detector row CT scanners. Precise segmentation of the lung nodules is critical, as it has a significant effect on subsequent analysis results \citep{macmahon2005guidelines}. However, for making an accurate diagnosis, a radiologist must check a CT scan containing about 150-500 slices, which is a very challenging and time-taking task \citep{way2010computer}. Moreover, it is hard to differentiate between the internal lung structure and the nodules, especially when the nodule is on the lung wall or attached to the end of a vessel in the lung tissue. The robust segmentation of lung nodules using simple threshold and morphological-based methods is very difficult \citep{kubota2011segmentation} due to the large variation in size and types of lung nodules ranging from adhesion-type nodules (juxtapleural and juxta-vascular) to GGO (ground-glass opacity) nodules. Fig.\ref{fig1} illustrates various types of lung nodules.

Another challenge in lung nodule segmentation is the segmentation of nodules with small diameter and intensity comparable to that of the surrounding noise. It is thereby hindering the downsampling potential of the segmentation network, where the network is unable to extract more in-depth network features that are semantic. This causes a significant impact on the accuracy of the extraction of feature maps of large nodules. For this reasons, a robust segmentation network is necessary to accommodate the large-scale nodule (various types) problem.

In the field of computer vision, convolutional neural networks (CNN) have recently become mainstream architecture. One such architecture, the U-Net, is an encoder-decoder like CNN architecture, which has shown exceptional results in the field of biomedical imaging on the task of segmentation \citep{ronneberger2015u}. Many modified U-Net architectures have achieved good results in different domains of biomedical imaging. However, CNN architectures implemented for the task of lung nodule segmentation are still immature. Therefore, the development of advanced architectures dealing with the shortcomings of previous architectures is essential. 

To deal with the challenges of efficient feature extraction and adaption to heterogeneity of lung nodules, a modified U-Net architecture with a weighted bidirectional feature network (U-Det) is proposed, which is appropriate for the segmentation of many forms of lung nodules. The following elements are the list of technical contributions through this research.
\begin{enumerate}
\item The proposed U-Det model uses a bidirectional feature network, which functions as a feature enricher, integrating multi-scale feature fusion for efficient feature extraction.
\item The implementation of data augmentation to deal with the relatively small size of the dataset and thereby preventing the model from over-fitting and generalizing better on the task of segmentation and providing robust performance.
\item The implementation of Mish activation function and class weights of masks to enhance model training and segmentation efficiency.
\item The U-Det model can achieve high segmentation performance on small nodules and various other challenging cases of nodule segmentation.
\end{enumerate}
\section{Background and related work}
\label{sec2}
This section describes several lung nodule segmentation strategies, like morphological methods, energy-based optimization techniques, region-growing processes, and machine learning methods that have been suggested in the past few years.

In morphological methods, morphological based-operations were applied for the task of removing the nodules attached to the vessels, and the isolation of lung nodules is done by selecting the connected region \citep{kostis2003three}. For improvement in separation of juxtapleural nodules from the lung wall, a morphological based-method combined with the shape hypothesis has been proposed for the replacement of the morphological arrangement of fixed-size \citep{sargent2017semi,kuhnigk2006morphological}. However, lung nodule segmentation using morphological operations is very challenging \citep{diciotti2011automated}.

In general, most region-growing methods are not capable of segmenting juxta-vascular and juxtapleural nodules and are just well suited to the isolation of calcified nodules \citep{kubota2011segmentation}. To address this issue, Dehmeshki et al. offered a region-growing operation that operates on information on intensity, fuzzy connectivity, distance, and peripheral contrast \citep{dehmeshki2008segmentation}. Here, a difficult task for these methods is the convergence condition. Also, irregular-shaped nodules are hard to process through region-growing methods due to the breach of the shape hypothesis.

The approach in energy based-optimization methods is to typically turn the task of segmentation into an energy minimization problem. For instance, in \citep{chan2001active,nithila2016segmentation,wang2016automatic,rebouccas2019new,farag2013novel}, the authors have proposed a level set function for the characterization of the image, and the energy function reaches the minimum, once the segmented contour meets the boundary of the lung nodule. Also, the maximum flow problem was used in \cite{ye2010automatic,boykov2004experimental,mukherjee2017lung}. Similar to the case of region-growing based operations, the presence of juxtapleural nodules and low contrast nodules such as GGO nodules drastically affects the output of these methods.

In the past decade, researchers in machine-learning field proposed hybrid models for classifying the lung nodules with high-level nodule segmentation feature maps \citep{lu2008accurate,lu2011effective}. In one such case, for the classification task, Lu et al. established a collection of features maps with the invariance of translation and rotation \citep{lu2013computer}. Another instance, Wu et al. proposed a culmination of texture and shape-dependent features for the classification of voxels, and a conditional random field (CRF) model was trained to classify voxels \citep{wu2010stratified}. Hu et al. performed the segmentation of the lungs and next carried out the Hessian matrix-based vascular feature extraction procedure to obtain the lung blood vessel mask. Then, the lung blood vessels were separated from the respective lung masks, and classification was achieved with the aid of a neural network \citep{hu2016neural}. Jung et al.'s method performs GGO nodular segmentation centered on multi-phase models, which are asymmetric and deformable \citep{jung2018ground}. \cite{gonccalves2016hessian} also developed a 3D large-scale nodule segmentation approach, based on the Hessian strategy.

Recently, to overcome the shortcomings of machine-learning based methods, Deep learning approaches have been proposed. In the deep learning approaches, CNN is a multi-layered neural network, which learns to map original image files and corresponding labels hierarchically, and the task of segmentation is modified into the classification of voxels similar to that used in the previous machine learning operations \citep{gao2016dropout,shen2017multi}. For instance, Wang et al. introduced the multi-view convolutional neural network (MVCNN) for nodule segmentation. The MVCNN is made up of three divisions of convolutional neural networks that are corresponding to three viewpoints of sagittal plane, axial plane, and coronal plane \citep{wang2017multi}. Zhao et al. have advocated an enhanced pyramid de-convolution network for enhanced performance on lung nodule segmentation. The architecture expertly blends low-level fine-grained characteristics with high-level functional characteristics \citep{zhao2019fine}. 

At the other end, Fully convolutional networks \citep{long2015fully} were a different method for the task of segmenting CT images. Few instances, \cite{ronneberger2015u} proposed 2D U-Net architecture and the 3D U-Net method advocated by \cite{cciccek20163d} are segmentation approaches which are better-adapted to biomedical imaging. Also, the central focused convolutional neural network(CF-CNN) proposed by Wang et al., which is a data-driven method without involving shape hypothesis, has shown strong performance for the segmentation of juxtapleural nodules \citep{wang2017central}. Recently, \cite{cao2020dual} proposed an approach of incorporating intensity features into the CNN architecture by implementing a Dual-branch residual network (DBResNet) to achieve attractive segmentation performance on lung nodules.

\section{Methods}
\label{sec3}
\begin{figure*}[!t]
\centering
\includegraphics[width=18cm,height=14cm,keepaspectratio]{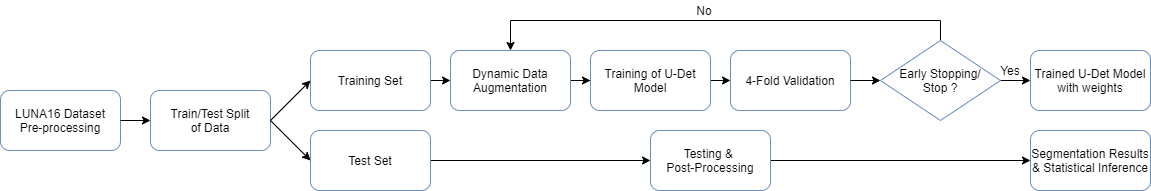}
\caption{Flow chart representing the overview of proposed model pipeline.}
\label{fig2}
\end{figure*}

This section covers the methods that are applied in proposed model. It consists of the following three phases: (1) Model Architecture, (2) Data Augmentation, and (3) Training and post-processing. Fig.\ref{fig2} illustrates the pipeline of proposed model.

\subsection{Model architecture}
\label{sec3_1}
\begin{figure*}[!t]
\centering
\includegraphics[width=18cm,height=14cm,keepaspectratio]{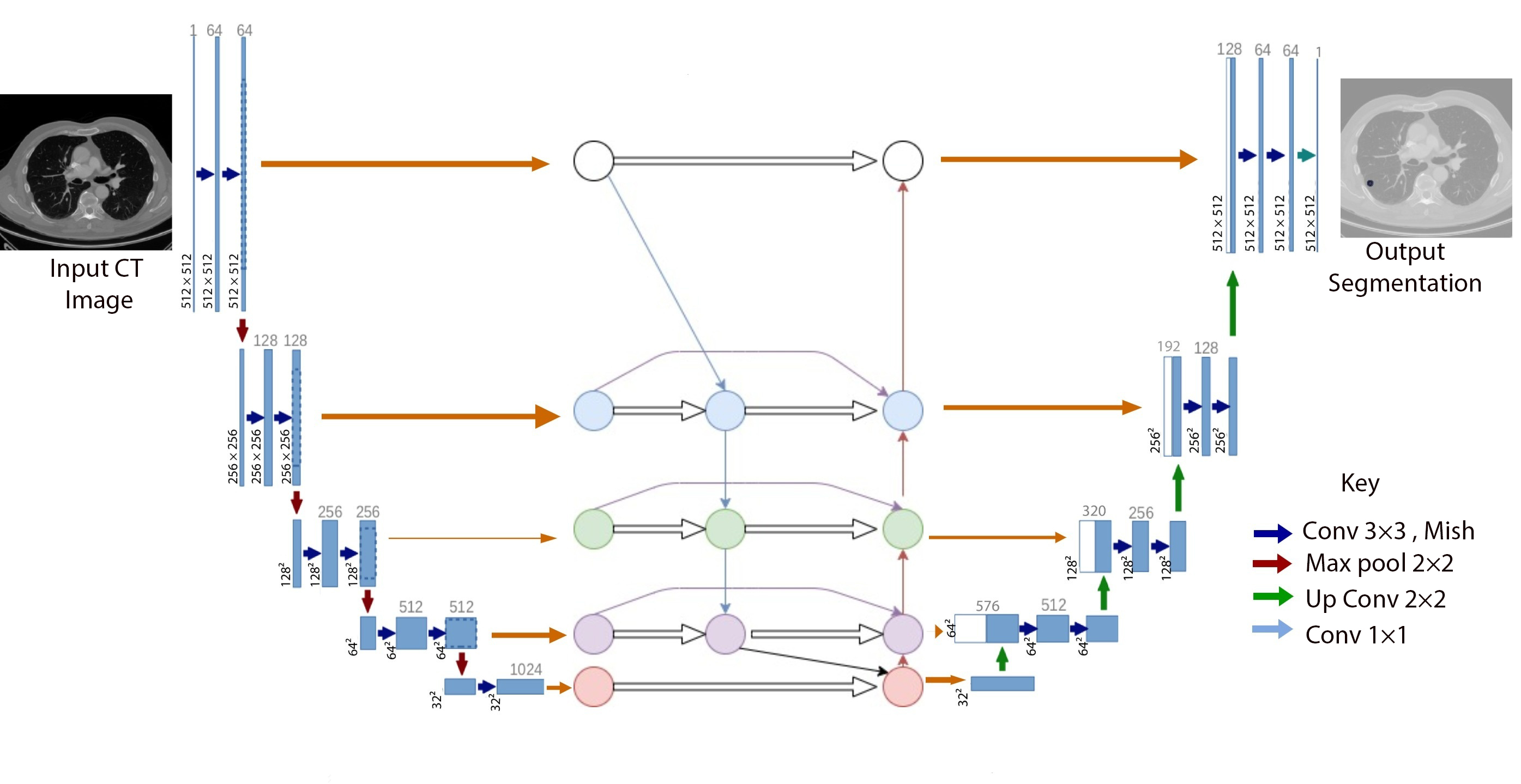}
\caption{Illustration of the proposed U-Det model, where the convolutional neural network block between the downsampling and up-sampling sections represents the Bi-FPN. The numbers at each layer of the architecture indicate the shape of feature maps at each layer, respectively. The key indicates the representation of various operations that take place in the backbone architecture.}
\label{fig3}
\end{figure*}

The model architecture is an End to End Deep learning approach for lung nodule segmentation. It takes inspiration from the encoder, decoder backbone of U-Net, and the feature enricher Bi-FPN, implemented in Efficient-Det. In this paper, the proposed model makes the use of U-Net based backbone network incorporated with a Bi-FPN for the task of lung nodule segmentation. Further, the fully convolutional network-based U-Net encoder takes the CT image, a slice of the CT scan, and outputs features at five corresponding depths, which are the respective inputs of the Bi-FPN. The feature network's outputs are combined respectively with a decoder architecture to obtain a combination of lower-level fine-grained features with high-level semantic features. The output mask represents lung nodules. Table.\ref{Table1} shows the corresponding layers of the model, along with their respective parameters. Fig.\ref{fig3} visualizes the proposed U-Det architecture.

\begin{table}
\centering
\caption{The layers and respective network parameters of the U-Det model.}
\label{Table1}
\begin{tabular}{@{}lc@{}}
\toprule
\textbf{Layer name}                & \multicolumn{1}{l}{\textbf{Number of parameters}} \\ \midrule
\textbf{Contraction path :}        & \multicolumn{1}{l}{}                              \\ \midrule
Conv2D$ \times $10, Mish           & 1.884$ \times 10^7$                                       \\ \midrule
MaxPool2D$ \times $4               &  -                             \\ \midrule
\textbf{BiFPN :}                   & \multicolumn{1}{l}{}                              \\ \midrule
Conv2D$ \times $5                  & 1.269$ \times 10^5$                                       \\ \midrule
BatchNormalization$ \times $12     & 3072                                              \\ \midrule
ReLU$ \times $12, MaxPool2D$ \times $3       & -                                                 \\ \midrule
DepthwiseConv$ \times $7                  & 4032                                              \\ \midrule
\textbf{Expansion path :}          & \multicolumn{1}{l}{}                              \\ \midrule
Conv2D$ \times $9, Mish                       & 6.821$ \times 10^6 $                                       \\ \midrule
Conv2DTrans$ \times $4, Mish                  & 2.786$ \times 10^6$                                       \\ \midrule
\textbf{Total parameters :}       & 2.858$ \times 10^7 $                                       \\ \bottomrule
\end{tabular}
\end{table}

\subsubsection{Backbone U-Net architecture}
\label{sec3_1_1}
The U-Net architecture is a convolutional network architecture for fast and precise segmentation of images. Recently, it has shown exceptional results in the field of biomedical image segmentation \citep{ronneberger2015u}. The proposed model uses a modified implementation of U-Net architecture to take a $512 \times 512$ image as an input and output a $512 \times 512$ mask. This architecture consists of two sections: the contraction and the expansion sections, which behave similarly to an encoder and decoder, respectively.

 In the architecture, the contracting path has the typical architecture of a convolutional network, and consists of repeated application of two $3 \times3$ convolutions (with `same' padding), each followed by a non-linear Mish activation function and a $2\times2$ max-pooling operation of stride 2 for downsampling of the input image features. The number of feature channels is doubled at each downsampling step. The depth of the contraction path is five. Also, for the regularization of the model, a Dropout layer with a dropout factor of 0.5 has been used after the second $3\times3$ convolution block at depth 4. The corresponding sizes of features at the five depths of the contraction section are $512\times512\times64$, $256\times256\times128$, $128\times128\times256$, $64\times64\times512$, $32\times32\times1024$ where  64, 128, 256, 512, 1024 represents the number of channels. 

The convolution process taking place at each layer of the model is denoted by the set of operations formulated below:
\begin{equation}
\label{Eq1}
    C[m,n] = (I  \times k)[m,n] =  \sum_{i} \sum_{j} k[i,j].I[m-i,n-j]
\end{equation}
\begin{equation}
\label{Eq2}
     Z^{[l]} = W^{[l]}.A^{[l-1]} + b^{[l]}
\end{equation}
\begin{equation}
\label{Eq3}
    A^{[l]} = f^{[l]}(Z^{[l]})
\end{equation}
where, Eq.\ref{Eq1} represents kernel convolution and Equations \ref{Eq2} and \ref{Eq3} denote the forward propagation process in CNN. Here in Eq.\ref{Eq1}, \textit{I} and \textit{k} denote the input image and kernel respectively. In Equations \ref{Eq2} and \ref{Eq3}, $ A^{[l]}, W^{[l]}, b^{[l]}, f^{[l]} $ indicate the activations, weights, bias, activation function of layer \textit{l} respectively.

The features at the five depths are input into the feature network (Bi-FPN), and the output feature vectors are input into the expansion section. Each step in the expansion path consists of an upsampling of the feature map followed by a $2\times2$ convolution (``up-convolution"), which halves the number of feature channels at each depth. The feature vectors obtained after upsampling are concatenated with the corresponding feature vectors from the feature network. The concatenation operation is followed by two $3\times3$ convolutions (`same' padding) and each followed by the Mish activation function. In the final layer of the backbone network, the obtained $512\times512\times64$ feature map undergoes two 3 $\times$ 3 convolutions. It is followed by the Mish activation function and a final $1\times1$ convolution block, and finally, sigmoid activation function. Thereby obtaining logits corresponding to the mask of the input CT image of the shape $512\times512$.

The network training aims to increase probability of right class of each voxel in the mask. To accomplish this, a weighted binary cross-entropy loss of each sample of training has been utilized. For the implementation of weighted binary cross-entropy, the positive pixels by the ratio of negative to positive voxels in the training set was weighted. Since the size of the positive class in a lung-nodule mask is relatively smaller than the size of the negative class, the class weight of the training set is positive thereby increasing the punishment for getting a positive value wrong. So the network will learn to be less biased towards outputting negative voxels due to the class imbalance in the masks. The weighted binary cross-entropy loss is formulated as follows:
\begin{equation}
\label{Eq4}
    Loss =  \frac{-1}{N}\sum_{i=1}^{ N } [ \omega _p \times  y_i  \log{ \widehat{y_i} + (1 -  y_i) \log{(1 - \widehat{y_i})}   } ]
\end{equation}
where, N represents the number of samples, $ \omega _p $ represents the positive prediction weights and $ \widehat{y_i} $ indicates the prediction of the U-Det model.

\subsubsection{Bi-FPN}
\label{sec3_1_2}
The Bi-FPN is based on the conventional top-down FPN (Feature Pyramid Networks) approach \citep{lin2017feature}. The Bi-FPN infuses efficient bidirectional cross-scale connections and weighted feature fusion into the model \citep{tan2019efficientdet}. Multi-scale feature fusion aims to fuse features at different resolutions to obtain efficient feature extractions. The one-way flow of information inherently limits conventional top-down FPN. A BiFPN does not consist of nodes that have only one input edge. If a node has only one input with no feature fusion, then it will contribute less to the feature network that aims to infuse different features. BiFPN also has one top-down and one bottom-up path, thereby allowing the bidirectional flow of features from one depth to the other in the feature network.

The incorporation of a bidirectional feature network aims to improve the feature extraction efficiency at each level of the backbone architecture and enrich the feature vectors, thereby allowing a fusion of lower-level fine-grained features and higher-level semantic features. As illustrated in Fig.\ref{fig3} the inputs of the Bi-FPN are the feature maps of the corresponding five depths of the contraction path of the backbone architecture. The outputs of Bi-FPN are fed into the expansion path of the backbone network.

 The BiFPN also incorporates additional weight for each input during feature fusion, thereby allowing the network to learn the particular input feature importance. For dynamic learning behavior and accuracy fast normalized fusion (one of the methods of incorporating weights during feature fusion) is implemented \citep{tan2019efficientdet}. Also, for improvement of efficiency, depthwise separable convolution followed by batch normalization and non-linear activation function ReLu (Rectified Linear unit) are implemented. Through the bidirectional cross-scale connections, the Bi-FPN enriches the feature maps at each depth of the network and provides an efficient fusion of features across various depths of the encoder section of the U-Net backbone architecture.
\subsubsection{Mish activation function}
\label{sec3_1_3}
In the neural network, the activation function is the gateway to incorporating nonlinearity. It plays a pivotal part in the training and evaluation of deep neural networks. The widely used activation functions are ReLU, Sigmoid, Leaky ReLU, Tan hyperbolic, and recently introduced Swish. The approach implements a recent state of the art activation function Mish, which works better than ReLU and Swish across challenging datasets. Furthermore, the simplicity of Mish makes it a smooth implementation in neural networks \citep{misra2019mish}.

\begin{figure}[!t]
\centering
\includegraphics[scale=.4]{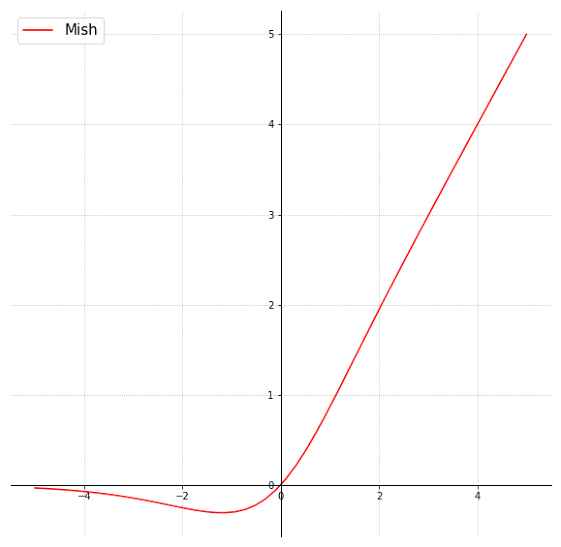}
\caption{The graphical plot of Mish activation function.}
\label{fig4}
\end{figure}

Mish is a non-monotonic and smooth neural network activation function formulated as:
\begin{equation}
\label{Eq5}
\centering
    f(x)=x.\tanh (\omega(x))
\end{equation}
where $ \omega(x)$ is the softplus activation function given by $ln(1 + e^x )$. Fig.\ref{fig4} illustrates the plot of the Mish activation function.

Mish implements a Self-Gating function, in which the input given to the gate is a scalar. The Self-Gating property helps replace activation functions such as ReLU (point-wise functions). Here, the input of the gating function is a scalar input with no requirement of modifying network parameters. In Tensorflow, the function definition of Mish is given by $x*tf.math.\tanh (tf.softplus(x))$. Mish's properties, like being above unbounded, below bounded, non-monotonic, and smooth, play a vital part in maximizing neural network outcomes. Hence, Mish enables considerable time improvements during the forward and backward pass on GPU (Graphics processing unit) inference, when CUDA (Compute Unified Device Architecture) is enabled, and improves the efficiency of the model.

\subsection{Data augmentation}
\label{sec3_2}
Medical image segmentation is constrained by the abundant availability of labeled training data. Data augmentation helps to prevent the model from over-fitting and helps in improving the generalization capability of the network on data outside the training set. It is vital in building robust deep learning pipelines \citep{mikolajczyk2018data,shorten2019survey}. In medical imaging, the augmentations are provided to both the image and label equally, thereby creating warped versions of the training data.

The number of annotated CT scans (lung nodules) in datasets is relatively less compared to other domains of application of Deep Learning. Thus an efficient implementation of sampling strategies or data augmentation is crucial for the robust performance of neural networks. Recently, Wang et al. and Cao et al., implemented weighted sampling of training data to deal with the comparatively smaller size of the LIDC-IDRI dataset \citep{wang2017central,cao2020dual}. In the above method, the input slice of the CT scan is cropped to a smaller size with the help of a random weighted sampling strategy to increase the size of the training dataset.

The proposed model inputs CT images of size $512\times512$, so a data augmentation strategy was followed instead of a sampling strategy to enhance the generalization potential and robustness of the proposed model. Data augmentation methods implemented in the proposed network are scale, flip, shift, rotate, and elastic deformations \citep{lalonde2018capsules}. Further, salt and pepper noise (impulse noise) was added to the input slice of the CT image to improve the generalizability of the proposed neural network. Along with the noise, elastic transformation, random shear, zoom, and rotation,  on the input image was implemented for maintaining the same input size. Thus by applying these small transformations to images during training, variety in the training dataset has been created and improved the robustness of the proposed U-Det model.

\subsection{Training and post-processing}
\label{sec3_3}
The training approach utilizes K-fold cross-validation \citep{bengio2004no} to obtain an accurate measure of the generalizing capability of the proposed model. To deal with the generation of augmented training CT images and corresponding ground truths, generators have been implemented for dynamic augmentation of input image and generation of corresponding ground truth labels. During model training, data augmentation and weighted binary cross-entropy deals with the data imbalance problem where the positive class was heavily over-weighed by the negative class in the masks.

In the training phase of proposed model, the `Adam'- model optimization algorithm \citep{kingma2014adam} was utilized with the following parameters : the initial learning rate is 0.0001, Beta\_1 = 0.99, Beta\_2 = 0.999, and decay rate is 1e-6. Also, a batch size of two samples was utilized to train the model. Further, the early stopping training strategy \citep{caruana2001overfitting} has been followed to prevent overfitting during the process of training the model.

In the post-processing phase, the proposed model has been designed to save the final obtained masks after the task of segmentation in a raw, metal (.mhd) format, which is one of the ways of storing volumetric data such as CT scans. Also, during testing, the proposed model has been designed to output qualitative figures, representing the final segmentation results and ground truth overlayed on the input CT image.

\section{Data and experiments}
\label{sec4}
This section deals with the data, implementation details, and assessment parameters (Evaluation metrics).
\subsection{Data}
\label{sec4_1}
\begin{figure}[!t]
\centering
\includegraphics[scale=.4]{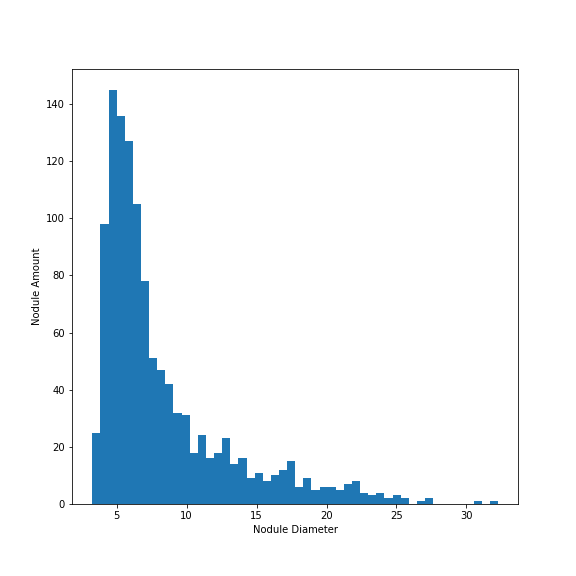}
\caption{Histogram of lung nodules size across the LUNA16 dataset.}
\label{fig5}
\end{figure}

For the experimentation and training of the proposed model, the approach utilizes the publicly available dataset of the Lung Nodule Analysis 2016 (LUNA16) grand challenge \citep{murphy2009large,jacobs2014automatic,setio2015automatic,van2009automatic}. This dataset is derived from the public dataset Lung Image Database Consortium and Image Database Resource Initiative (LIDC-IDRI) \citep{armato2011lung,setio2017validation}. It contains CT scans from the LIDC-IDRI database, where scans with a slice thickness 2.5 mm were excluded from the dataset. In total, 888 CT scans are included in the dataset. The LIDC-IDRI database contains annotations that were collected during a two-phase annotation process using four experienced radiologists. Here, the reference standard of the LUNA challenge consists of all nodules with a diameter greater than 3 mm accepted by at least three out of the four radiologists. The annotation file of the LUNA16 challenge contains annotations of 1186 nodules and also the enhanced annotations files indicating the various properties of the nodules. In Fig.\ref{fig5}, the histogram of nodule amount and diameter of nodules is illustrated .

After pre-processing, a total of 1166 CT images with corresponding ground truth masks were created and partitioned into two training and test subsets as 922 and 244, respectively. We applied K-fold cross-validation of 4-folds during training process. As depicted in Table.\ref{Table2}, the two subsets have identical statistical distribution in their clinical characteristics.

\begin{table}
\caption{Distribution of LUNA16 train and test sets. The values are indicated in the ``mean $ \pm $ standard deviation" format.} \vspace{0.08cm}
\label{Table2}
\centering
\begin{tabular}{lcc}
\hline
\textbf{Characteristics} & \textbf{\begin{tabular}[c]{@{}c@{}}Train Set\\ (n=922)\end{tabular}} & \textbf{\begin{tabular}[c]{@{}c@{}}Test Set\\ (n=244)\end{tabular}} \\ \hline
\textbf{Diameter(mm)}    & 8.13 $\pm $ 4.60                                                                        & 9.07 $ \pm $5.24                                                         \\ \hline
\textbf{Margin}          & 4.03 $\pm $ 0.82                                                                        & 4.06$\pm $ 0.76                                                         \\ \hline
\textbf{Spiculation}     & 1.60 $\pm $ 0.79                                                                        & 1.65 $\pm $ 0.87                                                         \\ \hline
\textbf{Lobulation}      & 1.73$\pm $ 0.73                                                                        & 1.82 $\pm $ 0.80                                                         \\ \hline
\textbf{Subtlety}        & 3.91 $\pm $ 0.82                                                                        & 4.06 $\pm $ 0.78                                                         \\ \hline
\textbf{Malignancy}      & 2.95 $\pm $ 0.92                                                                        & 3.03 $\pm $ 1.00                                                         \\ \hline
\end{tabular}
    \begin{tablenotes}
      \small
      \item Note: The range for all distinctive feature values except diameter is between 1 to 5. Moreover, the `margin' characteristic shows nodule edge clarity. `Spiculation' and `lobulation' indicate the shape characteristics of the nodule. `Subtlety' explains the contrast between the nodule zone and its surrounding areas. `Malignancy' reflects the possibility of this characteristic in a nodule.
    \end{tablenotes}
\end{table}

\subsection{Evaluation metrics}
\label{sec4_2}
The Dice similarity coefficient (DSC) is the key evaluation parameter for assessing the U-Det model's segmentation performance. It is a commonly used metric to calculate the difference between the outcomes of two segmentations \citep{valverde2017automated,havaei2017brain}. In addition to the above metrics, the sensitivity (SEN) and positive predictive value (PPV) have been used as auxiliary evaluation metrics. The evaluation metrics are formulated below:
\begin{equation}
\label{Eq6}
   DSC = \frac{2 \times V(Gt\cap Sv)}{V(Gt)+V(Sv)} 
\end{equation}    
\begin{equation}
\label{Eq7}
   SEN = \frac{V(Gt\cap Sv)}{V(Gt)} 
\end{equation}
 \begin{equation}
 \label{Eq8}
   PPV = \frac{V(Gt\cap Sv)}{V(Sv)} 
 \end{equation}
Where ``Gt" represents the ground truth labels, ``Sv" represents the segmentation results of the U-Det model. Here, volume size measured in voxel units is represented by `V.'
\subsection{Implementation details}
\label{sec4_3}
In this experiment, the Mish activation function (Section. \ref{sec3_1_3}) has been used for efficient training of the model, and also implementation of data augmentation was done on the LUNA16 training set to improve the robustness of the model (Section 3.2). Further, to prevent overfitting of the model, an early stopping training strategy was used; that is, if there is no more improvement in the performance of the model, then the model training will be stopped after an extra ten training epochs. Also, the strategy of reducing the learning rate of the optimizer on the model's performance reaching a plateau was implemented. The experiment is based on the Tensorflow (Version 2.1) deep learning framework (GPU version), and Python 3.6 is the language used for coding and also used CUDA 10.2 (for GPU computing) for accelerated training. The experiment was carried out on the google cloud platform on a virtual instance equipped with 4 vCPUs, 15GB memory, and an SSD drive of 500 GB. During the training of the model, acceleration was done on the NVIDIA Tesla T4 GPU (14 GB video memory), and it takes about 8 hours of training to converge. 
\section{Results and discussion}
\label{sec5}
This section covers the details of ablation study, overall performance of the proposed method, experimental comparison with other methods, and visualization of the results.

\subsection{Ablation Study}
\label{sec5_1}
An ablation experiment based on U-Net architecture has been designed. The ablation experiment verifies the effectiveness of each component in the proposed architecture. Table.\ref{Table3} shows the experimental results of the ablation study.

\subsubsection{Effect of Mish Activation Function}

In Table.\ref{Table3}, U-Net + Mish indicates the incorporation of the Mish activation function instead of ReLU activation function of original U-Net architecture. The DSC score of the original U-Net is 77.84 \%. After the implementation of the Mish function in U-Net, the DSC is observed to be 78.82 \%. Further, an encoder consisting of the contraction path of the U-Net along with the Bi-FPN, was implemented. On adding the Mish activation function to the above architecture, the DSC was 80.22 \%. Also, a version of the proposed U-Det model with ReLU is implemented, which performs marginally inferior to the mish version. It can be observed that the boost in performance due to Mish is nearly 1.3 \%. Thus it is evident that the Mish activation function is useful in the U-Det model.

\subsubsection{Effect of Bi-FPN}
 
In Table.\ref{Table3}, Encoder + Bi-FPN replaces the backbone U-Net architecture with only a contraction path and the Bi-FPN functioning as the feature enricher and decoder. It can be observed that this architecture shows improvement over the basic U-Net and achieves a DSC of 79.21 \%. Also, the ReLU version of the U-Det model is an incorporation of the Bi-FPN in the U-Net architecture, and it is observed that the DSC score is 81.63 \%, which is a significant improvement over the original U-Net. 
 
In addition to the above, even though the Bi-FPN is less computationally expensive than the expansive path of the U-Net architecture in terms of parameters, the Encoder + Bi-FPN successfully incorporates multiple features fusion. The multi-feature fusion thereby allows simultaneous feature map enhancement, thereby showing improvement over the U-Net architecture. Also, the multiple implementations of Bi-FPN may serve as a decoder pathway, but it results in more complexity, computational expense, and does not result in significant improvement. Thus it can be inferred that the implementation of Bi-FPN between the expansive and contractive paths is very effectual in the proposed model.

\subsubsection{Effect of Bi-FPN + Expansion path}
 
The combination of Bi-FPN and the expansion path (ReLU version of the U-Det model) has shown to be productive over the Encoder + Bi-FPN by exhibiting a DSC score of 81.63 \%. The addition of the expansion path of U-Net to the Encoder + Bi-FPN model helps in proper upsampling of low-level features and a combination of feature maps from Bi-FPN. Thereby enabling the efficient fusion of high-level semantic features with low-level features.

\subsubsection{Conclusion of the ablation study}
 
In Table.\ref{Table3}, on observation of the DSC score of the U-Det model (82.82 \%), it is evident that the proposed U-Det shows significant improvement over U-Net. Effectiveness of all components and their culmination in the proposed model is verified through the ablation study.

\begin{table*}[!ht]
\caption{Ablation Study on LUNA16 testing dataset. The study is based upon the U-Net model.}
\label{Table3}
\centering
\begin{tabular}{lllll}
\toprule
\textbf{Method}         & \textbf{DSC(\%)}     & \textbf{SEN(\%)}     & \textbf{PPV(\%)}     \\ \midrule
U-Net                   & 77.84 $ \pm $ 21.74          & 77.98 $ \pm $ 24.52          & 82.52 $ \pm $ 21.53          \\ \midrule
U-Net + Mish            & 78.82 $ \pm $ 22.01          & 78.97 $ \pm $ 24.83          & 83.56 $ \pm $ 21.80          \\ \midrule
Encoder + Bi-FPN        & 79.21 $ \pm $ 12.49          & 84.40 $ \pm $ 13.51          & 76.30 $ \pm $ 14.42          \\ \midrule
Encoder + Bi-FPN + Mish & 80.22 $ \pm $ 12.33          & 85.47 $ \pm $ 13.48          & 78.58 $ \pm $ 14.34          \\ \midrule
U-Det + ReLU            & 81.63 $ \pm $ 11.85          & 91.06 $ \pm $ 13.96          & 77.94 $ \pm $ 13.68          \\ \midrule
\textbf{U-Det}          & \textbf{82.82 $ \pm $ 11.71} & \textbf{92.24 $ \pm $ 14.14} & \textbf{78.92 $ \pm $ 17.52}
\end{tabular}
\end{table*}

\subsection{Overall performance}
\label{sec5_2}
\begin{figure}[!t]
\centering
\includegraphics[scale=.6]{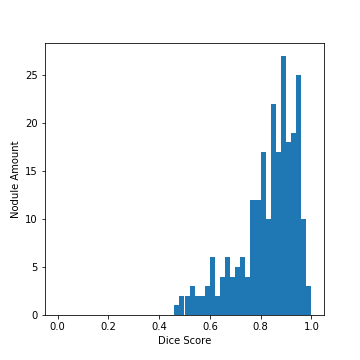}
\caption{LUNA16 testset DSC distributions.}
\label{fig6}
\end{figure}

The histogram of the DSC values and the total amount of nodules, centered on every sample in the test set, is plotted, as illustrated in Fig.\ref{fig6}, for better evaluation of the output of the U-Det model on test set. By observing Fig.\ref{fig6}, it can be quickly concluded that most nodules have a DSC value greater than 0.8.

\begin{figure*}[!ht]
\centering
\includegraphics[width=18cm,height=10cm,keepaspectratio]{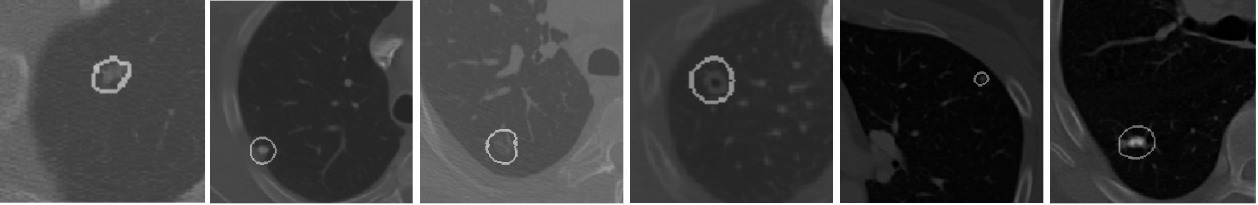}
\caption{Visualization of segmentation results of the proposed U-Det model on heterogeneous types of lung nodules. The various types of lung nodules from left to right: isolated nodule, juxtapleural nodule, a nodule of GGO and juxta-vascular type, cavitary nodule, nodule of very small size, calcific nodule.}
\label{fig7}
\end{figure*} 

For verification of the effectiveness of the Bi-FPN, the DSC results were compared with the original U-Net architecture. The U-Net model had a DSC of 77.84\%, whereas the proposed model has a DSC of 82.82\%, shows robust performance in the task of segmentation. So having a lesser number of parameters than the original U-Net architecture, the proposed U-Det model has shown its potential for efficient feature extraction and segmentation.

Further, the segmentation results of difficult cases including attached nodules (juxtapleural and juxta-vascular) and nodules of small size were studied. The mean DSC outcomes can be seen in Table.\ref{Table4}. By examining the experimental data shown in Table.\ref{Table4}, it is evident that the U-Det model's potential for robust segmentation is not dependent upon the type of nodule, and it performs exceptionally well on nodules of small size.

\begin{table*}
\caption{The segmentation results of the proposed model on various cases such as attached and non-attached nodules, and nodules of large and small sizes.}
\label{Table4}
\centering
\begin{tabular}{lllll}
\hline
                  & \multicolumn{4}{c}{\textbf{LUNA16 Test Set}}                                                                                                                                                                                                                                             \\ \hline
                  & \begin{tabular}[c]{@{}l@{}}Attached\\ (n=56)\end{tabular} & \begin{tabular}[c]{@{}l@{}}Non-Attached\\ (n=188)\end{tabular} & \begin{tabular}[c]{@{}l@{}}Diameter \textless 6 mm\\ (n=104)\end{tabular} & \begin{tabular}[c]{@{}l@{}}Diameter \textgreater{}= 6 mm\\ (n=140)\end{tabular} \\ \hline
\textbf{DSC (\%)} & 81.82                                                     & 83.11                                                          & 83.40                                                                     & 82.40    \\      \bottomrule                                               
\end{tabular}
\end{table*}

\subsection{Experimental comparison}
\label{sec5_3}
The results were compared with the results of other methods to depict the efficiency of the proposed method. The segmentation efficiency (DSC) of the four radiologists who worked on the LUNA16 (Derived from LIDC-IDRI) is known to be 82.25\%, and it can be observed that U-Det model performs better than the human experts. Also, the proposed U-Det model was compared with models ranging from the original U-Net to various other recent convolution networks, including the recent DB-ResNet. \citep{cao2020dual}.

In Table.\ref{Table5}, the quantified results of the various methods are represented. The outputs are in `` mean $ \pm $ standard deviation" format. As depicted in Table.\ref{Table5}, the U-Det model has shown better performance over the existing segmentation methods. Therefore, the U-Det model performs segmentation efficiently, having fewer parameters than the given models.

\begin{table*}[!ht]
\caption{The quantitative segmentation results of proposed model compared to different types of model architectures.}
\label{Table5}
\centering
\begin{tabular}{@{}llll@{}}
\toprule
\textbf{Network Architecture} & \multicolumn{1}{c}{\textbf{DSC (\%)}} & \multicolumn{1}{c}{\textbf{SEN (\%)}} & \multicolumn{1}{c}{\textbf{PPV (\%)}} \\ \midrule
FCN-UNET \citep{ronneberger2015u}                   & 77.84 $\pm $ 21.74                         & 77.98 $\pm $ 24.52                         & 82.52 $\pm $ 21.53                         \\ \midrule
CF-CNN \citep{wang2017central}                     & 78.55 $\pm $ 12.49                         & 86.01 $\pm $ 15.22                         & 75.79 $\pm $14.73                         \\ \midrule
MC-CNN \citep{shen2017multi}                     & 77.51 $\pm $ 11.4                          & 88.83 $\pm $ 12.34                         & 71.42 $\pm $ 14.78                         \\ \midrule
MV-CNN \citep{kang20173d}                    & 75.89 $\pm $ 12.99                         & 87.16 $\pm $ 12.91                         & 70.81 $\pm $ 17.57                         \\ \midrule
MV-DCNN \citep{wang2017multi}                & 77.85 $\pm $ 12.94                         & 86.96$\pm $ 15.73                         & 77.33 $\pm $ 13.26                         \\ \midrule
MCROI-CNN \citep{sun2017automatic}                   & 77.01 $\pm $ 12.93                         & 85.43 $\pm $ 15.97                         & 73.52 $\pm $ 14.62                         \\ \midrule
Cascaded-CNN \citep{havaei2017brain}              & 79.83 $\pm $ 10.91                         & 86.86$\pm $ 13.35                         & 76.14 $\pm $ 13.46                         \\ \midrule
DB-ResNet \citep{cao2020dual}                  & 82.74 $\pm $ 10.19                         & 89.35 $\pm $ 11.79                         & 79.64 $\pm $ 13.54                         \\ \midrule
\textbf{U-Det}                & \textbf{82.82 $\pm $ 11.71}                & \textbf{92.24 $\pm $ 14.14}                & \textbf{78.92 $\pm $ 17.52}                \\ \bottomrule
\end{tabular}
\end{table*}

\subsection{Visualization of results}
\label{sec5_4}

Even though MV-CNN, FCN U-Net, MCROI-CNN, MC-CNN, Cascaded-CNN and CF-CNN achieved good results, DB-ResNet shows better performance over them. Although the DB-ResNet achieves good performance in various cases, its performance is hindered in cases where the size of the nodule is less than 5mm \citep{cao2020dual} and in Fig.\ref{fig7}, the performance of the U-Det model on challenging cases such as small nodules, cavitary nodules, juxta-vascular, and juxtapleural nodules from the LUNA16 dataset is illustrated. By this observation it is evident that the proposed model U-Det has shown efficient performance on various types of nodules, including nodules of size less than 5mm.

\section{Conclusion}
\label{sec6}
This paper proposes an efficient modified U-Net architecture using a weighted bidirectional feature network (U-Det) for the segmentation of lung nodules. The model extracts and decodes feature maps through the backbone U-Net architecture, and the Bi-FPN acts as a feature enricher by incorporating multi-scale feature fusion. Through evaluation and visualization of the results of the proposed method, the proposed method demonstrated encouraging precision in the segmentation of the lung nodules and obtained an 82.82\% Dice similarity coefficient for the LUNA16 dataset. The U-Det model, in particular, successfully segments daunting cases such as cavitary nodules, GGO nodules, small nodules, and juxtapleural nodules. The future work focuses on developing a 3D Capsule Network based on the components of U-Det for fully automated malignancy classification of lung cancer.

\section*{Acknowledgments}
The authors acknowledge the LUNA16 grand challenge organizers, the National Cancer Institute, and the Foundation for the National Institutes of Health and their critical role in the creation of the publicly available LUNA16 Database for this study. We also acknowledge the Kaggle Data Science Bowl 2017 for providing insightful information regarding pre-processing.

\bibliographystyle{model2-names.bst}
\bibliography{Manuscript.bib}

\begin{thebibliography}{53}
\expandafter\ifx\csname natexlab\endcsname\relax\def\natexlab#1{#1}\fi
\providecommand{\url}[1]{\texttt{#1}}
\providecommand{\href}[2]{#2}
\providecommand{\path}[1]{#1}
\providecommand{\DOIprefix}{doi:}
\providecommand{\ArXivprefix}{arXiv:}
\providecommand{\URLprefix}{URL: }
\providecommand{\Pubmedprefix}{pmid:}
\providecommand{\doi}[1]{\href{http://dx.doi.org/#1}{\path{#1}}}
\providecommand{\Pubmed}[1]{\href{pmid:#1}{\path{#1}}}
\providecommand{\bibinfo}[2]{#2}
\ifx\xfnm\relax \def\xfnm[#1]{\unskip,\space#1}\fi
\bibitem[{Armato~III et~al.(2011)Armato~III, McLennan, Bidaut, McNitt-Gray,
  Meyer, Reeves, Zhao, Aberle, Henschke, Hoffman et~al.}]{armato2011lung}
\bibinfo{author}{Armato~III, S.G.}, \bibinfo{author}{McLennan, G.},
  \bibinfo{author}{Bidaut, L.}, \bibinfo{author}{McNitt-Gray, M.F.},
  \bibinfo{author}{Meyer, C.R.}, \bibinfo{author}{Reeves, A.P.},
  \bibinfo{author}{Zhao, B.}, \bibinfo{author}{Aberle, D.R.},
  \bibinfo{author}{Henschke, C.I.}, \bibinfo{author}{Hoffman, E.A.}, et~al.,
  \bibinfo{year}{2011}.
\newblock \bibinfo{title}{The lung image database consortium (lidc) and image
  database resource initiative (idri): a completed reference database of lung
  nodules on ct scans}.
\newblock \bibinfo{journal}{Medical physics} \bibinfo{volume}{38},
  \bibinfo{pages}{915--931}.
\bibitem[{Bengio and Grandvalet(2004)}]{bengio2004no}
\bibinfo{author}{Bengio, Y.}, \bibinfo{author}{Grandvalet, Y.},
  \bibinfo{year}{2004}.
\newblock \bibinfo{title}{No unbiased estimator of the variance of k-fold
  cross-validation}.
\newblock \bibinfo{journal}{Journal of machine learning research}
  \bibinfo{volume}{5}, \bibinfo{pages}{1089--1105}.
\bibitem[{Boykov and Kolmogorov(2004)}]{boykov2004experimental}
\bibinfo{author}{Boykov, Y.}, \bibinfo{author}{Kolmogorov, V.},
  \bibinfo{year}{2004}.
\newblock \bibinfo{title}{An experimental comparison of min-cut/max-flow
  algorithms for energy minimization in vision}.
\newblock \bibinfo{journal}{IEEE transactions on pattern analysis and machine
  intelligence} \bibinfo{volume}{26}, \bibinfo{pages}{1124--1137}.
\bibitem[{Cao et~al.(2020)Cao, Liu, Song, Hung, Ma, Xu, Jin and
  Lu}]{cao2020dual}
\bibinfo{author}{Cao, H.}, \bibinfo{author}{Liu, H.}, \bibinfo{author}{Song,
  E.}, \bibinfo{author}{Hung, C.C.}, \bibinfo{author}{Ma, G.},
  \bibinfo{author}{Xu, X.}, \bibinfo{author}{Jin, R.}, \bibinfo{author}{Lu,
  J.}, \bibinfo{year}{2020}.
\newblock \bibinfo{title}{Dual-branch residual network for lung nodule
  segmentation}.
\newblock \bibinfo{journal}{Applied Soft Computing} \bibinfo{volume}{86},
  \bibinfo{pages}{105934}.
\bibitem[{Caruana et~al.(2001)Caruana, Lawrence and
  Giles}]{caruana2001overfitting}
\bibinfo{author}{Caruana, R.}, \bibinfo{author}{Lawrence, S.},
  \bibinfo{author}{Giles, C.L.}, \bibinfo{year}{2001}.
\newblock \bibinfo{title}{Overfitting in neural nets: Backpropagation,
  conjugate gradient, and early stopping}, in: \bibinfo{booktitle}{Advances in
  neural information processing systems}, pp. \bibinfo{pages}{402--408}.
\bibitem[{Chan and Vese(2001)}]{chan2001active}
\bibinfo{author}{Chan, T.F.}, \bibinfo{author}{Vese, L.A.},
  \bibinfo{year}{2001}.
\newblock \bibinfo{title}{Active contours without edges}.
\newblock \bibinfo{journal}{IEEE Transactions on image processing}
  \bibinfo{volume}{10}, \bibinfo{pages}{266--277}.
\bibitem[{{\c{C}}i{\c{c}}ek et~al.(2016){\c{C}}i{\c{c}}ek, Abdulkadir,
  Lienkamp, Brox and Ronneberger}]{cciccek20163d}
\bibinfo{author}{{\c{C}}i{\c{c}}ek, {\"O}.}, \bibinfo{author}{Abdulkadir, A.},
  \bibinfo{author}{Lienkamp, S.S.}, \bibinfo{author}{Brox, T.},
  \bibinfo{author}{Ronneberger, O.}, \bibinfo{year}{2016}.
\newblock \bibinfo{title}{3d u-net: learning dense volumetric segmentation from
  sparse annotation}, in: \bibinfo{booktitle}{International conference on
  medical image computing and computer-assisted intervention},
  \bibinfo{organization}{Springer}. pp. \bibinfo{pages}{424--432}.
\bibitem[{Dehmeshki et~al.(2008)Dehmeshki, Amin, Valdivieso and
  Ye}]{dehmeshki2008segmentation}
\bibinfo{author}{Dehmeshki, J.}, \bibinfo{author}{Amin, H.},
  \bibinfo{author}{Valdivieso, M.}, \bibinfo{author}{Ye, X.},
  \bibinfo{year}{2008}.
\newblock \bibinfo{title}{Segmentation of pulmonary nodules in thoracic ct
  scans: a region growing approach}.
\newblock \bibinfo{journal}{IEEE transactions on medical imaging}
  \bibinfo{volume}{27}, \bibinfo{pages}{467--480}.
\bibitem[{Diciotti et~al.(2011)Diciotti, Lombardo, Falchini, Picozzi and
  Mascalchi}]{diciotti2011automated}
\bibinfo{author}{Diciotti, S.}, \bibinfo{author}{Lombardo, S.},
  \bibinfo{author}{Falchini, M.}, \bibinfo{author}{Picozzi, G.},
  \bibinfo{author}{Mascalchi, M.}, \bibinfo{year}{2011}.
\newblock \bibinfo{title}{Automated segmentation refinement of small lung
  nodules in ct scans by local shape analysis}.
\newblock \bibinfo{journal}{IEEE Transactions on Biomedical Engineering}
  \bibinfo{volume}{58}, \bibinfo{pages}{3418--3428}.
\bibitem[{El-Baz and Suri(2011)}]{el2011lung}
\bibinfo{author}{El-Baz, A.}, \bibinfo{author}{Suri, J.S.},
  \bibinfo{year}{2011}.
\newblock \bibinfo{title}{Lung imaging and computer aided diagnosis}.
\newblock \bibinfo{publisher}{CRC Press}.
\bibitem[{Farag et~al.(2013)Farag, El~Munim, Graham and Farag}]{farag2013novel}
\bibinfo{author}{Farag, A.A.}, \bibinfo{author}{El~Munim, H.E.A.},
  \bibinfo{author}{Graham, J.H.}, \bibinfo{author}{Farag, A.A.},
  \bibinfo{year}{2013}.
\newblock \bibinfo{title}{A novel approach for lung nodules segmentation in
  chest ct using level sets}.
\newblock \bibinfo{journal}{IEEE Transactions on Image Processing}
  \bibinfo{volume}{22}, \bibinfo{pages}{5202--5213}.
\bibitem[{Gao and Zhou(2016)}]{gao2016dropout}
\bibinfo{author}{Gao, W.}, \bibinfo{author}{Zhou, Z.H.}, \bibinfo{year}{2016}.
\newblock \bibinfo{title}{Dropout rademacher complexity of deep neural
  networks}.
\newblock \bibinfo{journal}{Science China Information Sciences}
  \bibinfo{volume}{59}, \bibinfo{pages}{072104}.
\bibitem[{Gon{\c{c}}alves et~al.(2016)Gon{\c{c}}alves, Novo and
  Campilho}]{gonccalves2016hessian}
\bibinfo{author}{Gon{\c{c}}alves, L.}, \bibinfo{author}{Novo, J.},
  \bibinfo{author}{Campilho, A.}, \bibinfo{year}{2016}.
\newblock \bibinfo{title}{Hessian based approaches for 3d lung nodule
  segmentation}.
\newblock \bibinfo{journal}{Expert Systems with Applications}
  \bibinfo{volume}{61}, \bibinfo{pages}{1--15}.
\bibitem[{Havaei et~al.(2017)Havaei, Davy, Warde-Farley, Biard, Courville,
  Bengio, Pal, Jodoin and Larochelle}]{havaei2017brain}
\bibinfo{author}{Havaei, M.}, \bibinfo{author}{Davy, A.},
  \bibinfo{author}{Warde-Farley, D.}, \bibinfo{author}{Biard, A.},
  \bibinfo{author}{Courville, A.}, \bibinfo{author}{Bengio, Y.},
  \bibinfo{author}{Pal, C.}, \bibinfo{author}{Jodoin, P.M.},
  \bibinfo{author}{Larochelle, H.}, \bibinfo{year}{2017}.
\newblock \bibinfo{title}{Brain tumor segmentation with deep neural networks}.
\newblock \bibinfo{journal}{Medical image analysis} \bibinfo{volume}{35},
  \bibinfo{pages}{18--31}.
\bibitem[{Hu and Menon(2016)}]{hu2016neural}
\bibinfo{author}{Hu, Y.}, \bibinfo{author}{Menon, P.G.}, \bibinfo{year}{2016}.
\newblock \bibinfo{title}{A neural network approach to lung nodule
  segmentation}, in: \bibinfo{booktitle}{Medical Imaging 2016: Image
  Processing}, \bibinfo{organization}{International Society for Optics and
  Photonics}. p. \bibinfo{pages}{97842O}.
\bibitem[{Jacobs et~al.(2014)Jacobs, van Rikxoort, Twellmann, Scholten,
  de~Jong, Kuhnigk, Oudkerk, de~Koning, Prokop, Schaefer-Prokop
  et~al.}]{jacobs2014automatic}
\bibinfo{author}{Jacobs, C.}, \bibinfo{author}{van Rikxoort, E.M.},
  \bibinfo{author}{Twellmann, T.}, \bibinfo{author}{Scholten, E.T.},
  \bibinfo{author}{de~Jong, P.A.}, \bibinfo{author}{Kuhnigk, J.M.},
  \bibinfo{author}{Oudkerk, M.}, \bibinfo{author}{de~Koning, H.J.},
  \bibinfo{author}{Prokop, M.}, \bibinfo{author}{Schaefer-Prokop, C.}, et~al.,
  \bibinfo{year}{2014}.
\newblock \bibinfo{title}{Automatic detection of subsolid pulmonary nodules in
  thoracic computed tomography images}.
\newblock \bibinfo{journal}{Medical image analysis} \bibinfo{volume}{18},
  \bibinfo{pages}{374--384}.
\bibitem[{Jung et~al.(2018)Jung, Hong and Goo}]{jung2018ground}
\bibinfo{author}{Jung, J.}, \bibinfo{author}{Hong, H.}, \bibinfo{author}{Goo,
  J.M.}, \bibinfo{year}{2018}.
\newblock \bibinfo{title}{Ground-glass nodule segmentation in chest ct images
  using asymmetric multi-phase deformable model and pulmonary vessel removal}.
\newblock \bibinfo{journal}{Computers in biology and medicine}
  \bibinfo{volume}{92}, \bibinfo{pages}{128--138}.
\bibitem[{Kang et~al.(2017)Kang, Liu, Hou and Zhang}]{kang20173d}
\bibinfo{author}{Kang, G.}, \bibinfo{author}{Liu, K.}, \bibinfo{author}{Hou,
  B.}, \bibinfo{author}{Zhang, N.}, \bibinfo{year}{2017}.
\newblock \bibinfo{title}{3d multi-view convolutional neural networks for lung
  nodule classification}.
\newblock \bibinfo{journal}{PloS one} \bibinfo{volume}{12}.
\bibitem[{Kingma and Ba(2014)}]{kingma2014adam}
\bibinfo{author}{Kingma, D.P.}, \bibinfo{author}{Ba, J.}, \bibinfo{year}{2014}.
\newblock \bibinfo{title}{Adam: A method for stochastic optimization}.
\newblock \bibinfo{journal}{arXiv preprint arXiv:1412.6980} .
\bibitem[{Kostis et~al.(2003)Kostis, Reeves, Yankelevitz and
  Henschke}]{kostis2003three}
\bibinfo{author}{Kostis, W.J.}, \bibinfo{author}{Reeves, A.P.},
  \bibinfo{author}{Yankelevitz, D.F.}, \bibinfo{author}{Henschke, C.I.},
  \bibinfo{year}{2003}.
\newblock \bibinfo{title}{Three-dimensional segmentation and growth-rate
  estimation of small pulmonary nodules in helical ct images}.
\newblock \bibinfo{journal}{IEEE transactions on medical imaging}
  \bibinfo{volume}{22}, \bibinfo{pages}{1259--1274}.
\bibitem[{Kubota et~al.(2011)Kubota, Jerebko, Dewan, Salganicoff and
  Krishnan}]{kubota2011segmentation}
\bibinfo{author}{Kubota, T.}, \bibinfo{author}{Jerebko, A.K.},
  \bibinfo{author}{Dewan, M.}, \bibinfo{author}{Salganicoff, M.},
  \bibinfo{author}{Krishnan, A.}, \bibinfo{year}{2011}.
\newblock \bibinfo{title}{Segmentation of pulmonary nodules of various
  densities with morphological approaches and convexity models}.
\newblock \bibinfo{journal}{Medical Image Analysis} \bibinfo{volume}{15},
  \bibinfo{pages}{133--154}.
\bibitem[{Kuhnigk et~al.(2006)Kuhnigk, Dicken, Bornemann, Bakai, Wormanns,
  Krass and Peitgen}]{kuhnigk2006morphological}
\bibinfo{author}{Kuhnigk, J.M.}, \bibinfo{author}{Dicken, V.},
  \bibinfo{author}{Bornemann, L.}, \bibinfo{author}{Bakai, A.},
  \bibinfo{author}{Wormanns, D.}, \bibinfo{author}{Krass, S.},
  \bibinfo{author}{Peitgen, H.O.}, \bibinfo{year}{2006}.
\newblock \bibinfo{title}{Morphological segmentation and partial volume
  analysis for volumetry of solid pulmonary lesions in thoracic ct scans}.
\newblock \bibinfo{journal}{IEEE Transactions on Medical Imaging}
  \bibinfo{volume}{25}, \bibinfo{pages}{417--434}.
\bibitem[{LaLonde and Bagci(2018)}]{lalonde2018capsules}
\bibinfo{author}{LaLonde, R.}, \bibinfo{author}{Bagci, U.},
  \bibinfo{year}{2018}.
\newblock \bibinfo{title}{Capsules for object segmentation}.
\newblock \bibinfo{journal}{arXiv preprint arXiv:1804.04241} .
\bibitem[{Lin et~al.(2017)Lin, Doll{\'a}r, Girshick, He, Hariharan and
  Belongie}]{lin2017feature}
\bibinfo{author}{Lin, T.Y.}, \bibinfo{author}{Doll{\'a}r, P.},
  \bibinfo{author}{Girshick, R.}, \bibinfo{author}{He, K.},
  \bibinfo{author}{Hariharan, B.}, \bibinfo{author}{Belongie, S.},
  \bibinfo{year}{2017}.
\newblock \bibinfo{title}{Feature pyramid networks for object detection}, in:
  \bibinfo{booktitle}{Proceedings of the IEEE conference on computer vision and
  pattern recognition}, pp. \bibinfo{pages}{2117--2125}.
\bibitem[{Long et~al.(2015)Long, Shelhamer and Darrell}]{long2015fully}
\bibinfo{author}{Long, J.}, \bibinfo{author}{Shelhamer, E.},
  \bibinfo{author}{Darrell, T.}, \bibinfo{year}{2015}.
\newblock \bibinfo{title}{Fully convolutional networks for semantic
  segmentation}, in: \bibinfo{booktitle}{Proceedings of the IEEE conference on
  computer vision and pattern recognition}, pp. \bibinfo{pages}{3431--3440}.
\bibitem[{Lu et~al.(2008)Lu, Barbu, Wolf, Liang, Salganicoff and
  Comaniciu}]{lu2008accurate}
\bibinfo{author}{Lu, L.}, \bibinfo{author}{Barbu, A.}, \bibinfo{author}{Wolf,
  M.}, \bibinfo{author}{Liang, J.}, \bibinfo{author}{Salganicoff, M.},
  \bibinfo{author}{Comaniciu, D.}, \bibinfo{year}{2008}.
\newblock \bibinfo{title}{Accurate polyp segmentation for 3d ct colongraphy
  using multi-staged probabilistic binary learning and compositional model},
  in: \bibinfo{booktitle}{2008 IEEE Conference on Computer Vision and Pattern
  Recognition}, \bibinfo{organization}{IEEE}. pp. \bibinfo{pages}{1--8}.
\bibitem[{Lu et~al.(2011)Lu, Bi, Wolf and Salganicoff}]{lu2011effective}
\bibinfo{author}{Lu, L.}, \bibinfo{author}{Bi, J.}, \bibinfo{author}{Wolf, M.},
  \bibinfo{author}{Salganicoff, M.}, \bibinfo{year}{2011}.
\newblock \bibinfo{title}{Effective 3d object detection and regression using
  probabilistic segmentation features in ct images}, in:
  \bibinfo{booktitle}{CVPR 2011}, \bibinfo{organization}{IEEE}. pp.
  \bibinfo{pages}{1049--1056}.
\bibitem[{Lu et~al.(2013)Lu, Devarakota, Vikal, Wu, Zheng and
  Wolf}]{lu2013computer}
\bibinfo{author}{Lu, L.}, \bibinfo{author}{Devarakota, P.},
  \bibinfo{author}{Vikal, S.}, \bibinfo{author}{Wu, D.},
  \bibinfo{author}{Zheng, Y.}, \bibinfo{author}{Wolf, M.},
  \bibinfo{year}{2013}.
\newblock \bibinfo{title}{Computer aided diagnosis using multilevel image
  features on large-scale evaluation}, in: \bibinfo{booktitle}{International
  MICCAI Workshop on Medical Computer Vision},
  \bibinfo{organization}{Springer}. pp. \bibinfo{pages}{161--174}.
\bibitem[{MacMahon et~al.(2005)MacMahon, Austin, Gamsu, Herold, Jett, Naidich,
  Patz~Jr and Swensen}]{macmahon2005guidelines}
\bibinfo{author}{MacMahon, H.}, \bibinfo{author}{Austin, J.H.},
  \bibinfo{author}{Gamsu, G.}, \bibinfo{author}{Herold, C.J.},
  \bibinfo{author}{Jett, J.R.}, \bibinfo{author}{Naidich, D.P.},
  \bibinfo{author}{Patz~Jr, E.F.}, \bibinfo{author}{Swensen, S.J.},
  \bibinfo{year}{2005}.
\newblock \bibinfo{title}{Guidelines for management of small pulmonary nodules
  detected on ct scans: a statement from the fleischner society}.
\newblock \bibinfo{journal}{Radiology} \bibinfo{volume}{237},
  \bibinfo{pages}{395--400}.
\bibitem[{Miko{\l}ajczyk and Grochowski(2018)}]{mikolajczyk2018data}
\bibinfo{author}{Miko{\l}ajczyk, A.}, \bibinfo{author}{Grochowski, M.},
  \bibinfo{year}{2018}.
\newblock \bibinfo{title}{Data augmentation for improving deep learning in
  image classification problem}, in: \bibinfo{booktitle}{2018 international
  interdisciplinary PhD workshop (IIPhDW)}, \bibinfo{organization}{IEEE}. pp.
  \bibinfo{pages}{117--122}.
\bibitem[{Misra(2019)}]{misra2019mish}
\bibinfo{author}{Misra, D.}, \bibinfo{year}{2019}.
\newblock \bibinfo{title}{Mish: A self regularized non-monotonic neural
  activation function}.
\newblock \bibinfo{journal}{arXiv preprint arXiv:1908.08681} .
\bibitem[{Mukherjee et~al.(2017)Mukherjee, Huang and
  Bhagalia}]{mukherjee2017lung}
\bibinfo{author}{Mukherjee, S.}, \bibinfo{author}{Huang, X.},
  \bibinfo{author}{Bhagalia, R.R.}, \bibinfo{year}{2017}.
\newblock \bibinfo{title}{Lung nodule segmentation using deep learned prior
  based graph cut}, in: \bibinfo{booktitle}{2017 IEEE 14th International
  Symposium on Biomedical Imaging (ISBI 2017)}, \bibinfo{organization}{IEEE}.
  pp. \bibinfo{pages}{1205--1208}.
\bibitem[{Murphy et~al.(2009)Murphy, van Ginneken, Schilham, De~Hoop, Gietema
  and Prokop}]{murphy2009large}
\bibinfo{author}{Murphy, K.}, \bibinfo{author}{van Ginneken, B.},
  \bibinfo{author}{Schilham, A.M.}, \bibinfo{author}{De~Hoop, B.},
  \bibinfo{author}{Gietema, H.}, \bibinfo{author}{Prokop, M.},
  \bibinfo{year}{2009}.
\newblock \bibinfo{title}{A large-scale evaluation of automatic pulmonary
  nodule detection in chest ct using local image features and
  k-nearest-neighbour classification}.
\newblock \bibinfo{journal}{Medical image analysis} \bibinfo{volume}{13},
  \bibinfo{pages}{757--770}.
\bibitem[{Nithila and Kumar(2016)}]{nithila2016segmentation}
\bibinfo{author}{Nithila, E.E.}, \bibinfo{author}{Kumar, S.},
  \bibinfo{year}{2016}.
\newblock \bibinfo{title}{Segmentation of lung nodule in ct data using active
  contour model and fuzzy c-mean clustering}.
\newblock \bibinfo{journal}{Alexandria Engineering Journal}
  \bibinfo{volume}{55}, \bibinfo{pages}{2583--2588}.
\bibitem[{Rebou{\c{c}}as~Filho et~al.(2019)Rebou{\c{c}}as~Filho,
  da~Silva~Barros, Almeida, Rodrigues and de~Albuquerque}]{rebouccas2019new}
\bibinfo{author}{Rebou{\c{c}}as~Filho, P.P.}, \bibinfo{author}{da~Silva~Barros,
  A.C.}, \bibinfo{author}{Almeida, J.S.}, \bibinfo{author}{Rodrigues, J.},
  \bibinfo{author}{de~Albuquerque, V.H.C.}, \bibinfo{year}{2019}.
\newblock \bibinfo{title}{A new effective and powerful medical image
  segmentation algorithm based on optimum path snakes}.
\newblock \bibinfo{journal}{Applied Soft Computing} \bibinfo{volume}{76},
  \bibinfo{pages}{649--670}.
\bibitem[{van Rikxoort et~al.(2009)van Rikxoort, de~Hoop, Viergever, Prokop and
  van Ginneken}]{van2009automatic}
\bibinfo{author}{van Rikxoort, E.M.}, \bibinfo{author}{de~Hoop, B.},
  \bibinfo{author}{Viergever, M.A.}, \bibinfo{author}{Prokop, M.},
  \bibinfo{author}{van Ginneken, B.}, \bibinfo{year}{2009}.
\newblock \bibinfo{title}{Automatic lung segmentation from thoracic computed
  tomography scans using a hybrid approach with error detection}.
\newblock \bibinfo{journal}{Medical physics} \bibinfo{volume}{36},
  \bibinfo{pages}{2934--2947}.
\bibitem[{Ronneberger et~al.(2015)Ronneberger, Fischer and
  Brox}]{ronneberger2015u}
\bibinfo{author}{Ronneberger, O.}, \bibinfo{author}{Fischer, P.},
  \bibinfo{author}{Brox, T.}, \bibinfo{year}{2015}.
\newblock \bibinfo{title}{U-net: Convolutional networks for biomedical image
  segmentation}, in: \bibinfo{booktitle}{International Conference on Medical
  image computing and computer-assisted intervention},
  \bibinfo{organization}{Springer}. pp. \bibinfo{pages}{234--241}.
\bibitem[{Sargent and Park(2017)}]{sargent2017semi}
\bibinfo{author}{Sargent, D.}, \bibinfo{author}{Park, S.Y.},
  \bibinfo{year}{2017}.
\newblock \bibinfo{title}{Semi-automatic 3d lung nodule segmentation in ct
  using dynamic programming}, in: \bibinfo{booktitle}{Medical Imaging 2017:
  Image Processing}, \bibinfo{organization}{International Society for Optics
  and Photonics}. p. \bibinfo{pages}{101332R}.
\bibitem[{Setio et~al.(2015)Setio, Jacobs, Gelderblom and van
  Ginneken}]{setio2015automatic}
\bibinfo{author}{Setio, A.A.}, \bibinfo{author}{Jacobs, C.},
  \bibinfo{author}{Gelderblom, J.}, \bibinfo{author}{van Ginneken, B.},
  \bibinfo{year}{2015}.
\newblock \bibinfo{title}{Automatic detection of large pulmonary solid nodules
  in thoracic ct images}.
\newblock \bibinfo{journal}{Medical physics} \bibinfo{volume}{42},
  \bibinfo{pages}{5642--5653}.
\bibitem[{Setio et~al.(2017)Setio, Traverso, De~Bel, Berens, van~den Bogaard,
  Cerello, Chen, Dou, Fantacci, Geurts et~al.}]{setio2017validation}
\bibinfo{author}{Setio, A.A.A.}, \bibinfo{author}{Traverso, A.},
  \bibinfo{author}{De~Bel, T.}, \bibinfo{author}{Berens, M.S.},
  \bibinfo{author}{van~den Bogaard, C.}, \bibinfo{author}{Cerello, P.},
  \bibinfo{author}{Chen, H.}, \bibinfo{author}{Dou, Q.},
  \bibinfo{author}{Fantacci, M.E.}, \bibinfo{author}{Geurts, B.}, et~al.,
  \bibinfo{year}{2017}.
\newblock \bibinfo{title}{Validation, comparison, and combination of algorithms
  for automatic detection of pulmonary nodules in computed tomography images:
  the luna16 challenge}.
\newblock \bibinfo{journal}{Medical image analysis} \bibinfo{volume}{42},
  \bibinfo{pages}{1--13}.
\bibitem[{Shen et~al.(2017)Shen, Zhou, Yang, Yu, Dong, Yang, Zang and
  Tian}]{shen2017multi}
\bibinfo{author}{Shen, W.}, \bibinfo{author}{Zhou, M.}, \bibinfo{author}{Yang,
  F.}, \bibinfo{author}{Yu, D.}, \bibinfo{author}{Dong, D.},
  \bibinfo{author}{Yang, C.}, \bibinfo{author}{Zang, Y.},
  \bibinfo{author}{Tian, J.}, \bibinfo{year}{2017}.
\newblock \bibinfo{title}{Multi-crop convolutional neural networks for lung
  nodule malignancy suspiciousness classification}.
\newblock \bibinfo{journal}{Pattern Recognition} \bibinfo{volume}{61},
  \bibinfo{pages}{663--673}.
\bibitem[{Shorten and Khoshgoftaar(2019)}]{shorten2019survey}
\bibinfo{author}{Shorten, C.}, \bibinfo{author}{Khoshgoftaar, T.M.},
  \bibinfo{year}{2019}.
\newblock \bibinfo{title}{A survey on image data augmentation for deep
  learning}.
\newblock \bibinfo{journal}{Journal of Big Data} \bibinfo{volume}{6},
  \bibinfo{pages}{60}.
\bibitem[{Siegel et~al.(2016)Siegel, Miller and Jemal}]{siegel2016cancer}
\bibinfo{author}{Siegel, R.L.}, \bibinfo{author}{Miller, K.D.},
  \bibinfo{author}{Jemal, A.}, \bibinfo{year}{2016}.
\newblock \bibinfo{title}{Cancer statistics, 2016}.
\newblock \bibinfo{journal}{CA: a cancer journal for clinicians}
  \bibinfo{volume}{66}, \bibinfo{pages}{7--30}.
\bibitem[{Sun et~al.(2017)Sun, Zheng and Qian}]{sun2017automatic}
\bibinfo{author}{Sun, W.}, \bibinfo{author}{Zheng, B.}, \bibinfo{author}{Qian,
  W.}, \bibinfo{year}{2017}.
\newblock \bibinfo{title}{Automatic feature learning using multichannel roi
  based on deep structured algorithms for computerized lung cancer diagnosis}.
\newblock \bibinfo{journal}{Computers in biology and medicine}
  \bibinfo{volume}{89}, \bibinfo{pages}{530--539}.
\bibitem[{Tan et~al.(2019)Tan, Pang and Le}]{tan2019efficientdet}
\bibinfo{author}{Tan, M.}, \bibinfo{author}{Pang, R.}, \bibinfo{author}{Le,
  Q.V.}, \bibinfo{year}{2019}.
\newblock \bibinfo{title}{Efficientdet: Scalable and efficient object
  detection}.
\newblock \bibinfo{journal}{arXiv preprint arXiv:1911.09070} .
\bibitem[{Valverde et~al.(2017)Valverde, Oliver, Roura, Gonz{\'a}lez-Vill{\`a},
  Pareto, Vilanova, Rami{\'o}-Torrent{\`a}, Rovira and
  Llad{\'o}}]{valverde2017automated}
\bibinfo{author}{Valverde, S.}, \bibinfo{author}{Oliver, A.},
  \bibinfo{author}{Roura, E.}, \bibinfo{author}{Gonz{\'a}lez-Vill{\`a}, S.},
  \bibinfo{author}{Pareto, D.}, \bibinfo{author}{Vilanova, J.C.},
  \bibinfo{author}{Rami{\'o}-Torrent{\`a}, L.}, \bibinfo{author}{Rovira,
  {\`A}.}, \bibinfo{author}{Llad{\'o}, X.}, \bibinfo{year}{2017}.
\newblock \bibinfo{title}{Automated tissue segmentation of mr brain images in
  the presence of white matter lesions}.
\newblock \bibinfo{journal}{Medical image analysis} \bibinfo{volume}{35},
  \bibinfo{pages}{446--457}.
\bibitem[{Wang and Guo(2016)}]{wang2016automatic}
\bibinfo{author}{Wang, J.}, \bibinfo{author}{Guo, H.}, \bibinfo{year}{2016}.
\newblock \bibinfo{title}{Automatic approach for lung segmentation with
  juxta-pleural nodules from thoracic ct based on contour tracing and
  correction}.
\newblock \bibinfo{journal}{Computational and mathematical methods in medicine}
  \bibinfo{volume}{2016}.
\bibitem[{Wang et~al.(2017a)Wang, Zhou, Gevaert, Tang, Dong, Liu and
  Tian}]{wang2017multi}
\bibinfo{author}{Wang, S.}, \bibinfo{author}{Zhou, M.},
  \bibinfo{author}{Gevaert, O.}, \bibinfo{author}{Tang, Z.},
  \bibinfo{author}{Dong, D.}, \bibinfo{author}{Liu, Z.}, \bibinfo{author}{Tian,
  J.}, \bibinfo{year}{2017}a.
\newblock \bibinfo{title}{A multi-view deep convolutional neural networks for
  lung nodule segmentation}, in: \bibinfo{booktitle}{2017 39th Annual
  International Conference of the IEEE Engineering in Medicine and Biology
  Society (EMBC)}, \bibinfo{organization}{IEEE}. pp.
  \bibinfo{pages}{1752--1755}.
\bibitem[{Wang et~al.(2017b)Wang, Zhou, Liu, Liu, Gu, Zang, Dong, Gevaert and
  Tian}]{wang2017central}
\bibinfo{author}{Wang, S.}, \bibinfo{author}{Zhou, M.}, \bibinfo{author}{Liu,
  Z.}, \bibinfo{author}{Liu, Z.}, \bibinfo{author}{Gu, D.},
  \bibinfo{author}{Zang, Y.}, \bibinfo{author}{Dong, D.},
  \bibinfo{author}{Gevaert, O.}, \bibinfo{author}{Tian, J.},
  \bibinfo{year}{2017}b.
\newblock \bibinfo{title}{Central focused convolutional neural networks:
  Developing a data-driven model for lung nodule segmentation}.
\newblock \bibinfo{journal}{Medical image analysis} \bibinfo{volume}{40},
  \bibinfo{pages}{172--183}.
\bibitem[{Way et~al.(2010)Way, Chan, Hadjiiski, Sahiner, Chughtai, Song,
  Poopat, Stojanovska, Frank, Attili et~al.}]{way2010computer}
\bibinfo{author}{Way, T.}, \bibinfo{author}{Chan, H.P.},
  \bibinfo{author}{Hadjiiski, L.}, \bibinfo{author}{Sahiner, B.},
  \bibinfo{author}{Chughtai, A.}, \bibinfo{author}{Song, T.K.},
  \bibinfo{author}{Poopat, C.}, \bibinfo{author}{Stojanovska, J.},
  \bibinfo{author}{Frank, L.}, \bibinfo{author}{Attili, A.}, et~al.,
  \bibinfo{year}{2010}.
\newblock \bibinfo{title}{Computer-aided diagnosis of lung nodules on ct
  scans:: Roc study of its effect on radiologists' performance}.
\newblock \bibinfo{journal}{Academic radiology} \bibinfo{volume}{17},
  \bibinfo{pages}{323--332}.
\bibitem[{Wu et~al.(2010)Wu, Lu, Bi, Shinagawa, Boyer, Krishnan and
  Salganicoff}]{wu2010stratified}
\bibinfo{author}{Wu, D.}, \bibinfo{author}{Lu, L.}, \bibinfo{author}{Bi, J.},
  \bibinfo{author}{Shinagawa, Y.}, \bibinfo{author}{Boyer, K.},
  \bibinfo{author}{Krishnan, A.}, \bibinfo{author}{Salganicoff, M.},
  \bibinfo{year}{2010}.
\newblock \bibinfo{title}{Stratified learning of local anatomical context for
  lung nodules in ct images}, in: \bibinfo{booktitle}{2010 IEEE Computer
  Society Conference on Computer Vision and Pattern Recognition},
  \bibinfo{organization}{IEEE}. pp. \bibinfo{pages}{2791--2798}.
\bibitem[{Ye et~al.(2010)Ye, Beddoe and Slabaugh}]{ye2010automatic}
\bibinfo{author}{Ye, X.}, \bibinfo{author}{Beddoe, G.},
  \bibinfo{author}{Slabaugh, G.}, \bibinfo{year}{2010}.
\newblock \bibinfo{title}{Automatic graph cut segmentation of lesions in ct
  using mean shift superpixels}.
\newblock \bibinfo{journal}{International journal of biomedical imaging}
  \bibinfo{volume}{2010}.
\bibitem[{Zhao et~al.(2019)Zhao, Sun, Qian, Qi, Sun, Zhang and
  Yang}]{zhao2019fine}
\bibinfo{author}{Zhao, X.}, \bibinfo{author}{Sun, W.}, \bibinfo{author}{Qian,
  W.}, \bibinfo{author}{Qi, S.}, \bibinfo{author}{Sun, J.},
  \bibinfo{author}{Zhang, B.}, \bibinfo{author}{Yang, Z.},
  \bibinfo{year}{2019}.
\newblock \bibinfo{title}{Fine-grained lung nodule segmentation with pyramid
  deconvolutional neural network}, in: \bibinfo{booktitle}{Medical Imaging
  2019: Computer-Aided Diagnosis}, \bibinfo{organization}{International Society
  for Optics and Photonics}. p. \bibinfo{pages}{109503S}.

\end{thebibliography}

\end{document}